# RealPen: Providing Realism in Handwriting Tasks on Touch Surfaces using Auditory-Tactile Feedback

**Youngjun Cho[1], Andrea Bianchi[2], Nicolai Marquardt[1] and Nadia Bianchi-Berthouze[1]**

[1]UCL Interaction Centre
Faculty of Brain Sciences
University College London, London, UK
[youngjun.cho.15, n.marquardt, n.berthouze]@ucl.ac.uk

[2]Industrial Design,
KAIST
Daejeon, South Korea
andrea@kaist.ac.kr

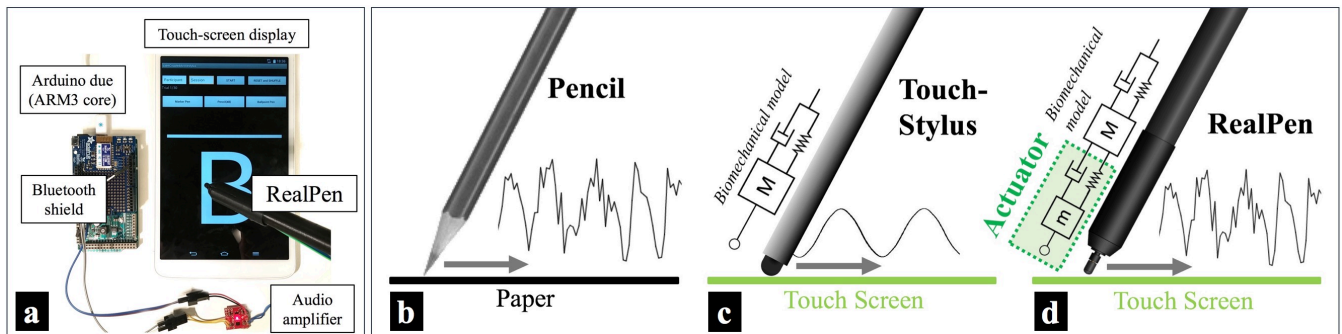

**Figure 1.** RealPen recreates the writing-sensations corresponding to different types of pens based on the analysis of friction-induced oscillations and sound of real pen tips moving on paper.

## ABSTRACT
We present RealPen, an augmented stylus for capacitive tablet screens that recreates the physical sensation of writing on paper with a pencil, ball-point pen or marker pen. The aim is to create a more engaging experience when writing on touch surfaces, such as screens of tablet computers. This is achieved by re-generating the friction-induced oscillation and sound of a real writing tool in contact with paper. To generate realistic tactile feedback, our algorithm analyses the frequency spectrum of the friction oscillation generated when writing with traditional tools, extracts principal frequencies, and uses the actuator's frequency response profile for an adjustment weighting function. We enhance the realism by providing the sound feedback aligned with the writing pressure and speed. Furthermore, we investigated the effects of superposition and fluctuation of several frequencies on human tactile perception, evaluated the performance of RealPen, and characterized users' perception and preference of each feedback type.

## ACM Classification Keywords
H.5.2 [information interfaces and presentation (e.g., HCI)]: User Interfaces – Tactile feedback, Auditory feedback.

## INTRODUCTION
Pen-based interaction with touch screens, such those found in commercially available tablet devices, is becoming increasingly popular. Recently, combining pen with touch-sensing allowed the design of novel expressive interaction techniques [21, 22]. But while graphical interfaces provide accurate simulation of diverse drawing tools (e.g., strokes of pens, pencils, brushes), the tactile sensation is usually limited to feeling a plastic pen tip sliding over a glass surface, limiting the overall experience when using digital drawing applications. Therefore, previous work has proposed various methods to address the lack of haptic and auditory feedback cues of the pen. For example, those techniques can provide feedback by using vibration motors [28], sound [29], or by generating textures for virtual objects [30, 35].

To provide a more realistic experience when interacting with a pen on digital surfaces, we investigate how to reproduce the *auditory-tactile* feelings of writing with physical drawing tools. High-definition haptic feedback is traditionally achieved by digitally generating textures rendered by means of motors [42] or electrical current [4]. Generally used for texture generation in touch interfaces these approaches do not simulate the haptic feedback due to the dynamic interaction between the surface and the object with which the surface is touched. Similarly auditory feedback has been widely explored with the main aim of creating engaging virtual sound effects [2, 19, 34] rather than providing a sound feedback for the simulated surface-object interaction. Our research investigates both high-definition haptic and auditory feedback for generating more realistic feedback for surface-pen interaction. Our aim is to simulate the rich multi-sensory feedback that is fundamental to drive and master our interaction with objects. Indeed, other than being possibly engaging, studies in neuroscience have shown that the multi-modal feedback expected from an action (e.g., in this case hand writing behavior with a specific pen over a specific type of texture) is critical to the sense of agency of the produced output of the action (e.g., the drawing) in the writer [39], drives our actions (e.g., amount of pressure while writing) and supports cognitive processes related to the action (e.g., learning to write or mastering the creation of drawing made possible by the drawing tools and surfaces [5, 12]).

In this paper, we present RealPen (Figure 1), which recreates the writing-sensations corresponding to different types of pens. Our approach is based on the analysis of the friction-induced oscillation and sound from real pen tips rubbed against the same type of paper (i.e., wood-pulp paper). The proposed methodology has two fundamental components: 1) tactile feedback is provided based on a combination of frequency fluctuation and superposition, and 2) auditory feedback is based on coupling effects of hand-writing speed and pressure on the selected type of paper. The paper is organized as follows: we introduce the concept of the RealPen and discuss related work. Next, we discuss the technical details for creating the tactile and auditory sensations. Then we report the results from two perceptual studies to characterize the performance and users' acceptance level of our prototype.

## REALPEN

RealPen is an augmented stylus that allows users to feel realism while writing and drawing on the glass surface of a capacitive touch-screen. Commercially available pens often use rubber-based capacitive material for the tip, which is capable of changing the capacitive field on the touch screen to get recognized as input. The friction force between the soft rubber material and the glass surface is subtle and different from the force between a traditional writing tool and the paper. The fundamental principle of RealPen is based on the analysis of different properties of the friction.

### Concept and Biomechanical Model

Figure 1d shows a schematic concept of the RealPen device. For traditional pencil and paper (Figure 1b), the dynamic friction coefficient is determined by the surface status of the paper and the graphite tip, and by the hand-writing behavior of the writer. Normally, the paper surface is composed of uneven and relatively complex texture structures (different to the glass material surface that is widely utilized as the top layer of tablet displays) and results in inducing dynamic and rough oscillations and sound when writing or drawing. The biomechanical model (1-DOF mass-spring-damper) of writing behavior with a pen on the touch display can be expressed as shown in Figure 1c. The hand gripping the pen can be expressed as the mass and the mechanical relationship between the tip and the touch surface decides the stiffness and the damping ratio. This basic model was presented by Schomaker and Plamondon [37] and was used to study how to graphically render pencil-drawings on virtual pallets or to identify writers [7, 17, 31]. Our key idea for creating the tactile illusions of the writing with the traditional writing tools is transforming the shape of original oscillations by adding an actuation part, performing an additional mass-spring-damper system (Figure 1d).

### Recreating of Realistic Tactile Feedback

Our algorithm for recreating realistic tactile pen feedback is shown in the flowchart of Figure 2. To generate driving signals for conveying tactile sensations, we first record the friction-induced oscillations using an accelerometer attached to a real analog pen, sampling at 1.344 kHz, and analyze the signal in the frequency domain after applying a high-pass filter of 30 Hz to remove any motion artifacts Figure 2a). The algorithm then extracts principal frequency components – the dominant frequency elements in terms of the power level – from each unit acceleration sequence (see

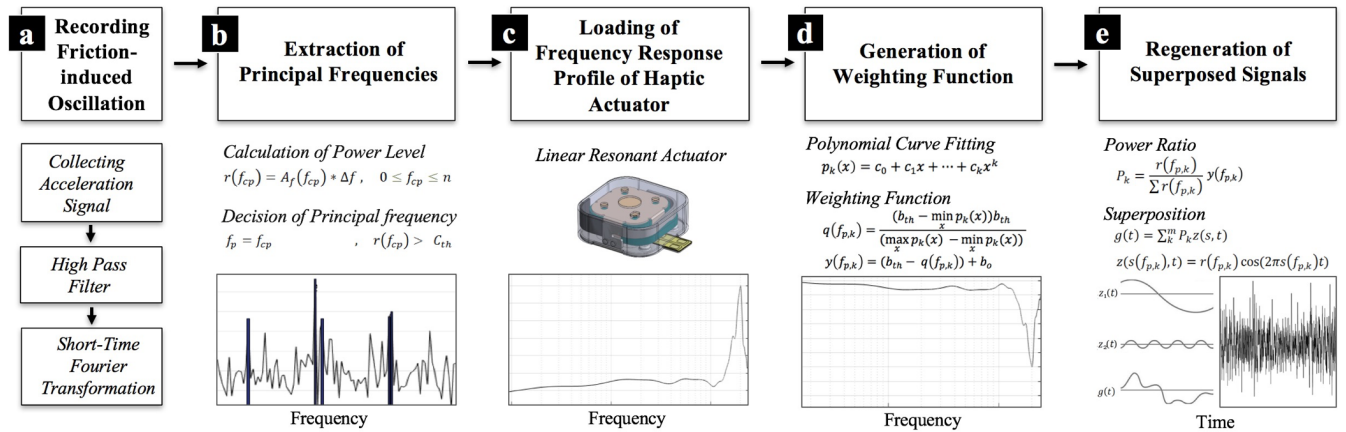

**Figure 2. RealPen algorithm for generating the tactile sensations of pen-writing.**

Figure 2b). The amplitude of the signal corresponding to each extracted frequency element from the unit sequence is adjusted by a weighting function to equalize the output strength (Figure 2d). This weighting function is calculated based on the frequency response profiles of the actuator (Figure 2c). All the extracted frequencies and the controlled amplitude are finally used for regenerating the superposed signals, depending on the power ratio (Figure 2e). The final signal patterns are then used by the actuation part of RealPen to deliver realistic tactile feedback.

**Coupling Effect on Sound: Pressure and Speed**
For the generation of auditory feedback we take a different strategy from the generation of tactile feedback. The main reason for this is the difference between the human audible frequency range (20 to 20,000Hz, [36, 45]) and the tactile sensitivity range of the mechanoreceptors (up to 500Hz, [15, 18, 40]) of normal healthy individuals. In addition, a wide range of commercialized audio devices allow us to capture and replay the sound signals. This is not the case for tactile signals. Therefore, to ensure the alignment of the two types of feedback, we couple the variances of the sound amplitude with the pressure and speed of hand-writing, instead of frequency properties. The same coupling phenomenon is indeed experienced when performing handwriting or drawing on paper in the physical world. In our prototype, the coupling is achieved by using the following linear mapping equation:

$$A_s = c_p(P_\omega - c_{op}) + c_x(\dot{x} - c_{ov}) \quad (1)$$

where $A_s$ is the amplitude of the auditory signal, $P_\omega$ is the writing pressure and $x$ is the writing displacement.

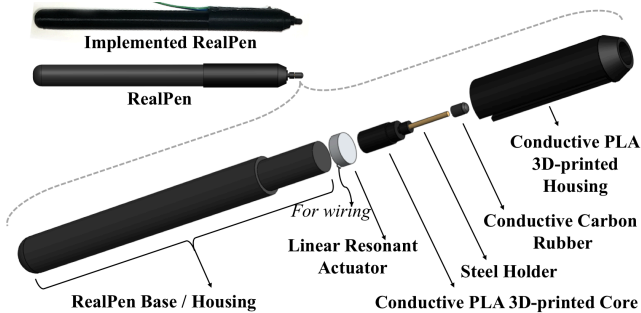

**Figure 3. Implemented RealPen and its schematic illustration.**

**Implementation**
We used a standard MakerBot 2X 3D printer to build the RealPen design by using both conductive (composite graphite PLA) and non-conductive materials (standard PLA). The schematic illustration of Figure 3 describes the implementation of RealPen (diameter: 11.5mm, length: 157mm)[1]. The device includes the conductive tip composed of carbon-based conductive rubber capable of stimulating capacitive fields on the touch surface. To convey the tactile sensation we utilize a linear resonant actuator (*LRA, Samsung Electro-mechanics*), a popular commercialized haptic actuators. The speaker embedded in the tablet device provides the sound feedback (see the bottom of Figure 6).

**Contribution**
The main contribution of RealPen is a new approach to generate auditory-tactile feedback that provides realism in hand-writing or drawing tasks on touch-based surfaces by concentrating on physical phenomena, friction-induced oscillation and sound. Our contributions are three-fold: 1) an algorithm for the generation of tactile sensations which finds principal frequency elements and reflects the physical properties of the haptic actuator, 2) the recreation of auditory-tactile feedback coupling for different types of pen-tips, and 3) two user evaluation studies of the effects of the above mechanisms on user experience during digital writing.

**RELATED WORK**
This section reviews the body of work of similar augmented pen-based interfaces that emulate the friction when writing on different materials, and the work about the synthesis of tactile textures.

**Friction for Traditional Writing Tools**
Friction is a significant part of a dynamic system. The friction force is coupled with oscillation and sound, and is entirely influenced by the types of materials of two rubbing surfaces [1]. During handwriting tasks, the friction force occurs between both surfaces (e.g. pencil – paper, stylus – touch-screen) so people sense the feeling of writing with their hand. Theoretically, the dynamic friction force induced by the moving object on the surface can be expressed as

$$F_s = -sgn(V)\,\mu_s F_n = m\ddot{x} \quad (2)$$

where $\mu_s$ is the dynamic friction coefficient and $x$ is the displacement. The dynamic friction force varies in accordance with the materials of the surface and the moving object. Note that the dynamic friction force excites vibrations and sounds, and the friction-excited vibrations and sounds play an intrinsic role in conveying feelings of handwriting.

Friction, oscillation, and pressure in human handwriting performance and their computational models are extensively discussed in literature [14, 24, 37]. For example, effects of pen-to-paper friction (i.e. ballpoint pen) when writing, and the relationship between pen force and the vertical displacement of pen tip are discussed by Schomaker et al. [37]. Dooijes et al. shows that the variations of writing pressure is accompanied with fluctuations in friction [14]. Chigira et al. analyzes the friction force between a ballpoint pen and paper in relation to handwriting [10]. Although previous work highlights the effects of friction on handwriting with traditional tools, these works have not shown how to render the friction related phenomena. Furthermore, little effort has been spent on the analysis of the friction-induced oscillation and sound by means of a digitally superimposed signals activated by software.

---
[1] Our STL (STereoLithography) files for 3D-printing a RealPen design are available at:
http://youngjuncho.com/index.php/2016/uist-2016-realpen/.

**Pen with Auditory-Haptic Feedback**
Similar to visual feedback, auditory and haptic feedback influence human perception in both the real and the virtual world. Research on auditory-haptic feedback has shown promise in enhancing realism of traditional unimodal interfaces. For example, haptic feedback has been used with sound feedback in interacting with media contents [8, 11], for alert notifications [23, 32], and in pen-based applications [13, 29, 34, 38]. In previous work on pen-based displays, auditory feedback has been used for the delivery of verbal information [29], notification sounds as a confirmation method [13, 29], and nonvisual guidance [2, 34, 38]. As for the haptic feedback, haptic devices can be classified into two categories; a) tactile (i.e., felt by mechanoreceptors and b) kinesthetic (i.e., awareness of movement and force generated by the muscles) devices [20]. The same classification can be applied to pen-based haptic interfaces. For example, pen interfaces provide tactile confirmation by means of vibrations [28, 29], textures [30, 35] and flow-rotation effects [3]. On the other end, Kamuro et al. [26], Chen et al. [9] and Withana et al. [44] present pen-based tangible techniques which utilize kinesthetic feedback. Work on pen-based passive haptic feedback [43] can be also included in this category. Extending this earlier work, our aim is to bring realism of physicality to handwriting with digital interfaces. We focus on the properties of friction-induced oscillation and sound (i.e. auditory-tactile) in handwriting, while the kinesthetic cue is beyond the scope of the work discussed in this paper.

**Techniques for Texture Generation and Rendering**
In order to create virtual textures, a wide range of approaches have been exploited. These span from utilizing psychophysical phenomena like electro-vibration [4] and squeeze effect [6], to presenting localized actuation based on additional haptic interfaces like pin-array devices [25] or interactive pen displays [30, 35]. These techniques allow new multi-dimensional interaction by conveying feelings in the virtual world. The work presented in this paper is inspired by the techniques above mentioned, specifically those related with pen-writing behavior (e.g., friction-induced oscillation and sound) with different types of pen-tips. Specifically, in this paper we consider the physical friction produced between the moving objects (i.e. pen-tips) and the stationary surfaces rather than rendering the feelings of the virtual surface and touched objects [30, 35]. So for example, the property (i.e. frequency response) of the haptic actuator should also be considered in the process of texture creation in order to keep the specific frequency bands from dominating the tactile perception.

## TECHNICAL DETAILS FOR TACTILE SENSATIONS
We use this section to describe the technical details of the RealPen algorithm for generating the realistic tactile sensations. First, we address the principle of the extraction of principal frequency elements. Then, we discuss the need of reflecting the actuator's physical characteristics, and the method to generate the tactile patterns. Last, we describe the hardware composition to activate our algorithm.

**Extraction of Principal Frequencies**
The original oscillation signals measured from the real pen rubbed against paper (see Figure 6, top) are analyzed using a short-time Fourier transform. Based on the calculation of the power level on the frequency spectrum of each unit sequence (see Figure 4a), principal frequencies can be determined as shown in Figure 4b. Here, we then need to select two important parameters: 1) the number of principal frequencies for frequency superposition; 2) the size of the unit sequence which is matched with that of the sub-pattern of the final tactile sensation. Finding the optimal settings for these parameters is the goal of our first user study.

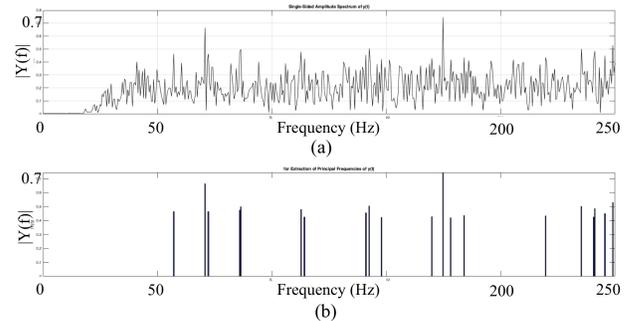

**Figure 4.** a) FFT result, b) extracted principal frequencies.

**Actuator and Reflecting Frequency Response Profile**
There are a wide variety of miniaturized actuators which can be used in hand-held interfaces: e.g. eccentric rotating mass motor, linear resonant actuator, or piezo actuators. To recreate the rich tactile sensations, it is crucial to select the actuator capable of being easily controlled in terms of the expression of frequency elements. In the case of an eccentric rotating mass motor, for example, it creates concentrated force commensurate with the square of the number of the motor's revolution, that is, we cannot change its frequency without altering its magnitude. Therefore, we decided to use linear actuators for RealPen, such as linear resonant actuators and piezoelectricity-based actuators.

Previous work [4, 27] determined that when the linear actuator is driven at a periodic electrical signal, the mechanical displacement of the actuating mass is influenced by frequency. Generally, it peaks to around its resonant frequency and dramatically decreases beyond the frequency (e.g. the graph in Figure 2c). Therefore, the tactile rendering algorithm must be capable of reflecting the frequency response profile of the actuator. The final output should have equalized strength within the whole range of the working frequency, going beyond earlier research on texture creation [30, 35]. The equalization process in our proposed algorithm can be applied to any type of the linear actuator including voice-coil based actuators, whose underlying mechanical principle is similar to that of linear resonant actuators in terms of Q-factor selection [27].

**Generation of Tactile Patterns**
To generate the final tactile pattern, our method is based on superposition of principal frequencies (see the left of Figure 5). The pattern can be composed of either multiple sub-patterns or only one. The number of sub-patterns is related to the term of frequency fluctuation. We will discuss this in the next section when presenting the user study on human tactile perception. For each sub-pattern, we use the extracted principal frequency from a unit sequence in original signals, and generate sinusoidal waves as many as the number of the extracted frequency elements. Then, all the waves of each frequency component are superposed together reflecting both the weighting function and each power level of the corresponding frequency element. The total length of the final tactile pattern is 1500 milliseconds (Figure 5, right graph) and this is continuously repeated while the person is writing or drawing with the pen.

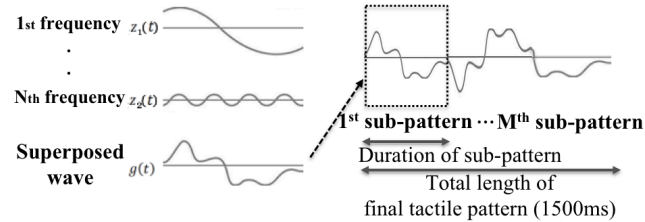

Figure 5. Key factors: superposition (left) and frequency fluctuation (right).

**Hardware Design**
Figure 6 shows the hardware composition for implementing our method. The top of the Figure depicts the hardware composition for recording and analyzing the original oscillations from a real pen rubbed against the surface. The tactile patterns produced by the algorithm are programmed in the micro-processor (Atsam3x8e, Atmel) as shown by the flow in the right of Figure 6. To drive the actuator (LRA, Samsung Electro-Mechanics), the tactile digital patterns are analog-converted and amplified as soon as being activated by the wireless communication through the Bluetooth module (Ez-Link Bluefruit) connected to the processor when the screen of the tablet (LG G-pad) is touched (Figure 6, bottom row).

## STUDY 1: HUMAN TACTILE PERCEPTION OF REALPEN
To further investigate optimal technical parameters for the RealPen hardware as well as getting insights into people's perception of the tactile feedback we conducted two user studies. In this section, we report the procedure and results of our first user study evaluating the perception of tactile properties produced by RealPen. The results are then used to finalize the RealPen technical parameters to be used in the subsequent studies. For this first experiment we focus on tactile feedback perception, and therefore did not include any auditory feedback.

**Key factors: Frequency Superposition and Fluctuation**
Frequency superposition and fluctuation, depicted in Figure 5, are our main parameters for the study on human tactile perception. Previous works [4, 40] on human psychophysics related to the tactile perception have reported psychological responses to stimuli (with specific frequencies and strength) in terms of the sensations and the just-noticeable-differences (JND). However, few researchers have focused on the effects of multiple frequencies (i.e. frequency superposition) and those variances (i.e. fluctuation) on human perception. Considering the fact that the human mechanoreceptors can feel complex oscillations like the friction-induced vibrations in writing tasks [16, 40], our user study explored how those two factors impact on people's perception.

For our study, we designed six conditions with two independent variables, the number of *principal frequencies* (i.e. superposition) and the number of *sub-patterns* (i.e. fluctuation). The three conditions for the superposition are: the top-five, the top-ten, and the top-fifteen principal frequencies. In parallel, two conditions for the fluctuation are: the fifteen sub-patterns and the only one pattern (see Figure 5). Accordingly, we have a total of 3x2 = 6 conditions as described in Table 1.

In this study, we would like to answer these questions: *How much do the frequency fluctuation and the superposition have an impact on realism? Will users feel comfortable when provided with the tactile sensations while writing with the pen? What parameters are most suitable for recreating*

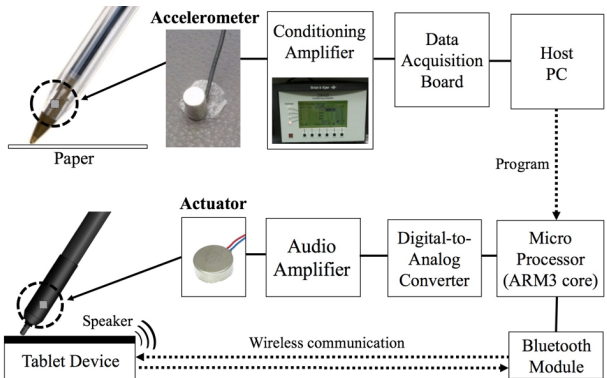

Figure 6. Hardware of RealPen setup.

| Condition | A | B | C | D | E | F |
|---|---|---|---|---|---|---|
| The number of principal frequencies (Superposition) | 5 | 10 | 15 | 5 | 10 | 15 |
| # of sub-patterns / each duration (Fluctuation) | 15 / 100 ms | 15 / 100 ms | 15 / 100 ms | 1 / 1500 ms | 1 / 1500 ms | 1 / 1500 ms |

Table 1. Six conditions for the first user study to investigate the human tactile perception.

| Questions | Treatment | Block | $F_r$ | Df | p |
|---|---|---|---|---|---|
| Q1) Does it feel like writing with a real pen (i.e. ballpoint pen)? | Superposition | Fluctuation | 6.45 | 2 | 0.040 |
| | Fluctuation | Superposition | 8.91 | 1 | 0.003 |
| Q2) Is it comfortable to write? | Superposition | Fluctuation | 5.66 | 2 | 0.059 |
| | Fluctuation | Superposition | 2.88 | 1 | 0.090 |

**Table 2. Results using Friedman's two-way ANOVA by rank.**

the sensation?

**Procedure and Participants**
We recruited 12 healthy adults (7 female) of varying ethnicity (aged 24-41 years, M=29.5, SD=4.92) from the local university and non-research community. Each subject was given the information sheet and the informed consent form prior to data acquisition. The experiment was conducted in a quiet lab room with no distractions and took approximatively half an hour per participant. The study was conducted following a within-group design, where each participant tested the interface for each of three conditions of frequency superposition (top-5, 10, 15) with two different numbers of sub-patterns (15, 1). All the conditions were counterbalanced in Latin squared design.

All participants were asked to fill out a questionnaire that consisted of demographics and six sessions corresponding to each condition. During every session, each participant was asked to perform one common task: drawing lines and writing a letter 'B' twice on the tablet device (LG Gpad) with the RealPen. After each task, the participants was asked to rate on a five point Likert scale *(ranging from 1, not at all, to 5, very much)* the following two questions: 1) *Does it feel like writing with a real pen (i.e. ballpoint pen)?*, and 2) *Is it comfortable to write?*. After finishing the final session, we collected descriptions of their writing experience in their own words as feedback. The experimental protocol was approved by the Ethics Committee of University College London Interaction Centre.

**Software Design**
An Android-based paint application was designed for this study. To remove effects of visual feedback on human perception, the color of drawing strokes was set to have the same color of the background palette (i.e, black).

**Results**
The collected data was analyzed using Friedman's two-way analysis of variance (ANOVA) with alpha value 0.05. Table 2 provides a summary of the results of the statistical test. In addition, Figure 7 summarizes the result of the first user study with 95% confidence interval.

The results show a significant effect of both frequency superposition ($F_r$=6.45, p<0.05) and fluctuation ($F_r$=8.91, p<0.005) on how realistically the users perceived the tactile feedback. As we expected, both factors are hence critical parameters to generate realistic tactile sensations through the pen-to-paper writing.

In the case of comfort (Question type 2), there is no significant difference among each independent variable (the number of superposed frequencies, the number of sub-patterns, respectively), although *condition C (top 15 principal frequencies and 15 sub-patterns)* was the highest rated among all conditions (M= 3.667, SD=0.651).

In order to decide the final value for each parameter, we finally used the nonparametric Wilcoxon signed rank test for post hoc analysis of paired conditions about the ratings of question 1 (see Table 2). When comparing within the same subset of pairs with the same fluctuation value, only conditions D and F (see Table 1) showed a significant difference between their ratings (p=0.025). When comparing between conditions with the same number of frequency components but different fluctuation value, only condition A vs D does not show any difference (B-E: p=0.046, C-F: p=0.047). The results indicate that the fluctuated tactile signal (i.e., A,B,C) provides users with a more realistic feeling of writing than the non-fluctuated signal when the top-ten or fifteen principal frequency elements are superposed. On the other hand, the number of superposed frequencies over ten had no significant impact on the realism of writing when having fluctuated signals. This fact provided a foundation for designing efficient and effective tactile patterns which are used with auditory feedback to convey realism in handwriting tasks. Indeed, there is a limitation on processing resources, which are required to superpose different waves, so the findings are more valuable.

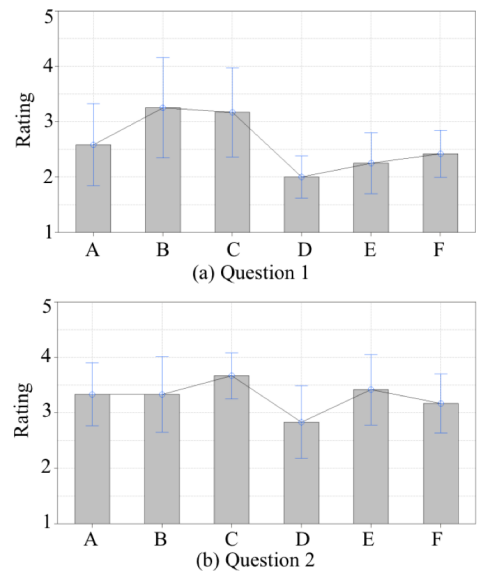

**Figure 7. Plot of 95% confidence interval in ratings of realism and comfort (X axis: Condition – see table 1, Y axis: Likert scale);** *Q1. Does it feel like writing with a real pen? , Q2. Is it comfortable to write?*

Participants rated their sensations when given the *condition B (top 10 principal frequencies and 15 sub-patterns)* on average as 3.250 (SD=1.422) on a five-point Likert scale as most realistic (see Figure 7a). Therefore, we decided to use values corresponding to the condition B for the recreation of the final tactile patterns.

## RECREATION OF AUDITORY-TACTILE SENSATIONS

In this section, we describe the design of the pen final tactile cues, incorporating the results of our previous tactile perception study and adding auditory feedback.

### Target Auditory-Tactile Sensations

We reviewed a wide range of commercialized paint applications for the iOS and Android platforms (e.g. QMemo by LG, SketchBook by Autodesk, Sketch by Sony), and identified common writing and drawing tools, such as a ballpoint pens, marker pens, pencil or brushes. The three types of tools listed in Table 3 were selected and modelled in RealPen. Notebook paper was selected as the only surface for our experiments. This decision was made to keep the study focused on the pen and to keep the number of conditions small.

| Pen Tip | Writing Surface |
|---|---|
| Ballpoint Pen | Notebook (Papers) |
| Pencil (4B) | |
| Marker Pen | |

**Table 3. RealPen differentiates feelings of several types of pens/pencils rubbed against sheets of paper.**

### Tactile Feedback

We measured the raw friction-induced oscillation signals from the three types of pens rubbed against a paper surface with natural writing speed and pressure. Considering the characteristics of our algorithm equalizing the output strength over the active frequency range of the linear actuator, the processing efficiency is important due to limited resources. In this respect, the identified value for the superposition (i.e., top-10 principal frequencies) is crucial to most effectively create the required tactile patterns. Subsequently, the top-ten frequency elements were extracted from each pen every 100 milliseconds. To collect the array of extracted frequencies recorded consistently without motion artifacts and errors, we threw away the first and last recordings and picked out 15 continuous sets (i.e. 1500 milliseconds signals) of 10 frequency elements from the middle. Our RealPen algorithm processes each extracted frequency set to superpose the tactile signals corresponding to each pen as depicted in the left side of Figure 8. Each pair of graphs includes the time-domain signal and the frequency-domain signals using a short-time Fourier transformation. The completed patterns are repeated in a loop while users write with the RealPen.

### Auditory Feedback

The auditory feedback of RealPen is based on the coupling effect of writing pressure and speed. We first recorded raw friction sound from the three types of pens, similar to the process used for the measurement of the tactile oscillations. For each sound, we scanned the audio signal with a 300 milliseconds-window and picked out a unit sequence which has a similar pattern and amplitude in the beginning and the end. With this procedure we made sure that the sound can be played smoothly when the 300 ms auditory signal is continuously repeated. To reflect the coupling effect of pressure and speed, equation (1) was applied to the final auditory feedback. The tablet device depicted in Figure 6 was used to calculate relative pressure of writing and writing-velocity and to play the final auditory feedback. Friction sounds of the pencil and marker have similar frequency spectrums as shown in the right of Figure 8. For conducting the following user study, the maximum amplitude of each auditory pattern was equalized to minimize the impact of the volume of the sound on the users' performance to distinguish auditory feedbacks of each pen.

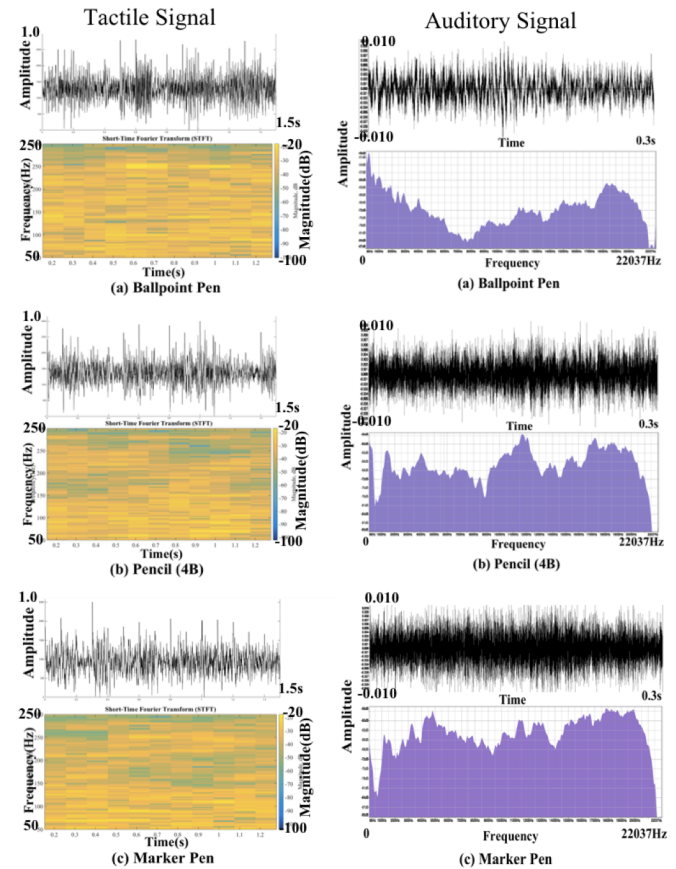

**Figure 8. The final patterns: tactile signals (left) and auditory signals (right) in the time and frequency domain.**

## STUDY 2: AUDITORY-TACTILE PERCEPTION OF REALPEN

We conducted a second user study to investigate people's perception of auditory-tactile properties of RealPen.

### Experimental Setup and Procedures

The experiment consisted of two parts: 1) a recognition test to evaluate the auditory-tactile feedback provided by RealPen in terms of the error rate and the response time, and 2) an auditory-tactile perception study to investigate how people think and feel about each feedback modality of the RealPen. Figure 9 describes the experimental set up for the study. An android-based application was designed to accurately collect the user's behaviour as well as reported perception of the RealPen feedback. The user behavioural measures were: users' selections of the recognized feeling, response time (i.e., time duration required to match the perceived sensations), and trajectory profiles including writing speed, and pressure. To block the effect of handedness, the graphical user interface was designed in bilateral symmetry (see Figure 9). The menu bars for the operator were hidden during the experiment.

Part 1 of the experiment, the recognition test, had 30 fully randomized trials, including 12 training trials (3 types of pens x 4 repetitions) used as the exposing period and 18 experimental trials (3 types of pens x 6 repetitions) for each modality of the three conditions: a) tactile feedback, b) auditory feedback, and c) auditory-tactile feedback. All conditions were counterbalanced. Each participant was first given a physical pen, marker and pencil to feel the auditory and tactile feelings of the real physical tool while drawing several lines on a physical piece of paper. Then, each subject was instructed to draw a horizontal line and write the letter 'B', which is composed by both straight and curved lines, using our RealPen device up to a maximum of three times before matching the presented feelings with one of the auditory-tactile sensations. We limited to the relatively less complex task to manage any possible negative influence and boredom.

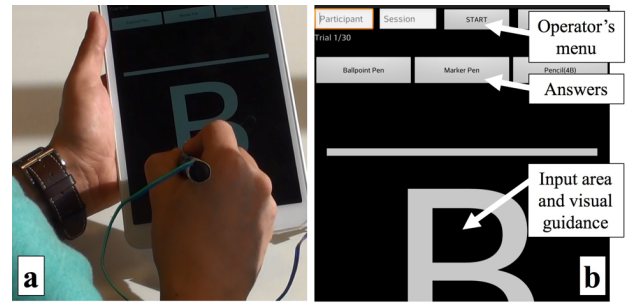

Figure 9. (a) Experimental set up used to test the auditory-tactile perception and (b) the GUI for the first part of the study.

In the second part of the test, the auditory-tactile perception study, each subject was instructed to perform free drawings containing zigzag lines and circles with the RealPen to answer the self-reported questionnaires. In this part of the experiment, they were given three auditory-tactile sensations corresponding to the ballpoint pen only (selected as a representative tool). Participants rated each feedback modality according to the Likert scale based questions: *Does it feel like writing with the real pen (i.e. Ballpoint Pen)? (1: not at all, 5: very much)*, *Is it comfortable to write? (1: not at all, 5: very much)*, *Is the sensation ...? (1: too subtle, 5: too strong)*, *Is the auditory-tactile feedback well balanced?*

### Participants

To remove learning effects, we invited 12 new participants (8 female) who had never experienced RealPen before the experiment. Their age ranged from 18 to 43 (M=29.83, SD=7.09), and one participant was left-handed. Like the tactile perception study, the information sheet, consent and demographic forms were given before beginning the experiment. Each participant executed the 30 trials for each of the three feedback conditions in the first part, and free drawings in the second part. The total experiment time for each subject was between 50 and 70 minutes.

### Results

648 trial samples were collected (12 participants x 18 experimental trials x 3 different feedback types) and were tested using two-way repeated measures ANOVAs with the pairwise Bonferroni adjustment for post-hoc analysis.

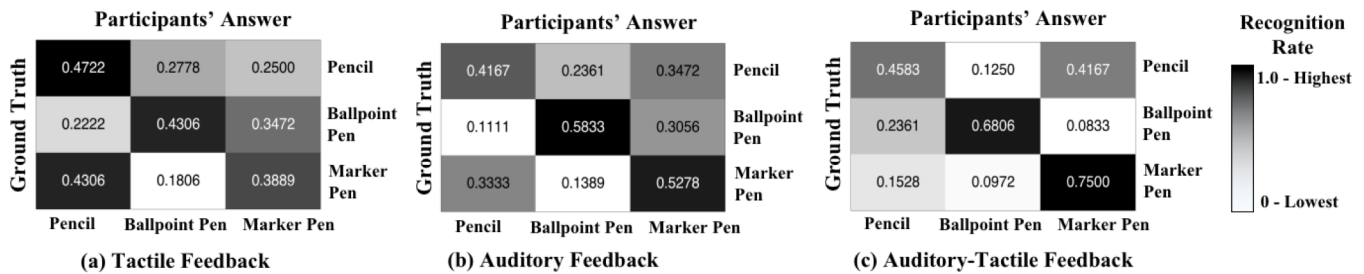

Figure 10. Results of the Recognition Test using Confusion Matrices: (a) with tactile feedback, (b) with auditory feedback, (c) with auditory-tactile feedback.

| Source | Sum of Square | Df | Mean Square | F | Significance |
|---|---|---|---|---|---|
| Modality (Type of Feedback) | 4.343 | 2 | 2.171 | 9.473 | 0.000 |
| Error (Modality) | 32.546 | 142 | 0.229 | | |
| Type of Pen | 1.787 | 2 | 0.894 | 3.245 | 0.042 |
| Error (Type of Pen) | 39.102 | 142 | 0.275 | | |
| Modality * Type of Feedback | 2.843 | 4 | 0.711 | 3.962 | 0.004 |
| Error (Modality * Type of Feedback) | 50.935 | 284 | 0.179 | | |

**Table 4. Results of the recognition rate using two-way repeated measures ANOVA**

The recognition rate for the different modalities is summarized in Figure 10. Overall, the total rates of each modality, tactile, auditory, and auditory-tactile feedback, are 43.05%, 50.96%, and 62.96%, respectively. The recognition rate showed a significant major effect of modality (type of feedback) ($F(2, 142)=9.473$, $p<0.001$), and a significant effect of the type of tools ($F(2, 142)=3.245$, $p<0.05$). As we expected, participants performed differently in accordance with the modality, and they also showed the different recognition rate for each type of tools. In addition, the interaction between the two independent variables was also significant ($F(4,284)=3.962$, $p<0.005$), indicating that the partcipants could better recognize the tool when both the auditory and tactile feedback were combined. The results of the statistical analysis are summarized in Table 4.

As for the pairwise comparisons, significant differences were found between tactile feedback and auditory-tactile feedback ($p<0.001$), and between auditory feedback and auditory-tactile feedback ($p<0.005$). However, no differences were found between the two unimodal forms of feedback ($p=0.443$). With the modality being fixed, no significant differences were found between pair of the pen types.

During the trial experiments we also measured the response time, that is how long it took for the participants to match the perceived sensations to the writing tools. Figure 11 shows the response time of each condition with 95% confidence interval. The shortest average response time was observed with the auditory-tactile feedback (Tactile: M=9.247s, SD=5.086s; Auditory: M=8.851s, SD=6.498s; Auditory-Tactile: M=8.497s, SD=4.406s). However, there was no significant effect of the type of pen and the different feedback modality on the response time with respectively $F(2,142)=0.860$ ($p=0.425$) and $F(2,142)=1.927$ ($p=0.149$). We found no significant interaction effect between modality and pen type ($F(4,284)=0.798$, $p=0.527$).

Finally, we analysed the ratings of the experience of the 12 participants using a non-parametric Friedman's test. Participants rated the auditory-tactile feedback of the RealPen as more realistic than other conditions (M=3.75, SD=0.965, on 5-point Likert scale) (see Figure 12). Indeed, the statistical analysis showed an effect of feedback modality on the perceived realism ($\chi^2= 8.211$, df=2, $p<0.05$). However, no effect of modality on comfort was found (Tactile: M=3.583, SD=0.9; Auditory: M=3.333, SD=0.888; Auditory-Tactile: M=3.417, SD=0.669; $\chi^2=1.077$, df=2, $p=0.584$). From the question asking strength, we were not able to find any significant differences ($\chi^2= 3.257$, df=2, $p=0.196$). Nonetheless, users responded that each feedback was well balanced but it felt stronger when the combined feedback (i.e., auditory and tactile) was given (M=3.5, SD=0.798) in comparison to unimodal feedback (Tactile: M=3.0, SD=0.739; Auditory: M=3.25, SD=0.965). This suggests that the strength of each form of feedback should be adjusted for recreating the tactile sensations.

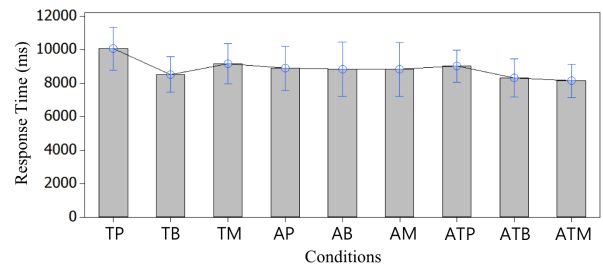

**Figure 11. Plot of 95% confidence interval in the response time; Conditions: TP(Tactile, Pencil), TB(Tactile, Ballpoint pen), TM(Tactile, Marker pen), AP(Auditory, Pencil), AB(Auditory, Ballpoint pen), AM(Auditory, Marker pen), ATP(Auditory-Tactile, Pencil), ATB(Auditory-Tactile, Ballpoint pen), ATM(Auditory-Tactile, Marker pen).**

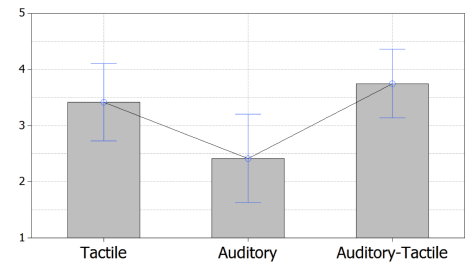

**Figure 12. Plot of 95% confidence interval in ratings of realism for the ballpoint pen (X axis: modality, Y axis: Likert scale).**

### Discussion

The effects of multimodal feedback have been investigated in previous research for different applications [8, 33, 41]. Our own study expands beyond this earlier research and we contribute the comparison of auditory feedback, tactile feedback and auditory-tactile feedback in terms of realism in handwriting tasks. In particular, we aimed to answer the fundamental questions of: *"Do multiple frequency components and those variances have influences on human perception.", "Will a combination of tactile and auditory* *feedback affect human perception and recognition performance?"*, and *"Will we in turn bring realism of physicality into the handwriting with digital interfaces?"*

We targeted to recreate the feelings of the three types of pen, which are common for everyday writing tasks (i.e., the pencil, ballpoint pen, and marker pen). Both the tactile and auditory signals from marker and pencil are very similar to each other, whereas the measured signals of the ballpoint pen are more unique in terms of their frequency spectrum as analyzed above (see the section *Recreation of auditory-tactile feeling* and Figure 9). As we expected, people have more difficulty distinguishing the marker and the pencil under each unimodal feedback (the average recognition rates were 43.06% and 47.23% for tactile feedback and auditory feedback, respectively). However, the recognition rate under the auditory-tactile feedback was substantially improved up to 60.42% on average. Overall, participants showed better performance (over 60% recognition rate) when the multimodal feedback were provided in comparison with the unimodal sensations. This result was also accompanied with higher satisfaction in regard to realism from the followed self-reported questionnaires. As addressed above, we only focus on the tactile feedback when rendering the haptic sensations. To reach full realism and improve the overall scores, kinesthetic feedback should also be considered in future iterations of this work - in fact, both tactile and kinesthetic sensations are equivalently fundamental to haptic manipulation [20], and play an important role in handwriting tasks [16]. The kinesthetic feedback could be realized in the pen-based interfaces by using additional multiple degree-of-freedom force feedback or by controlling the viscosity of the pen-tip using a magnetorheological (MR) fluid.

Lastly, a growing body of literature in psychology and in HCI highlights the importance of rich multimodal feedback to support cognitive processes in various context where digital writing is used: from kids learning to write [12] to supporting artists in the creation of digital artworks [5]. We expect that by providing the full range of physical feedback, users will be able to make full use of the writing surfaces and writing tools and learn in a natural way how to fully exploit the potential of such tools.

**CONCLUSION**

RealPen is a new interactive technology that enables a touch pen to provide auditory-tactile illusions of pen-and-paper writing on a touch surface based on the recreation of friction-induced oscillation and sound of different types of pens. Two separate user studies were conducted to explore human perception of RealPen, with study results on effects of frequency fluctuation and superposition on realism related to handwriting and comparing realism of each feedback modality. We believe this advanced feedback approach for pen-based input has the potential to be extended in further research about supporting people's interaction with tablet computers and investigating human psychophysics.